\documentclass{elsart}
\usepackage{epsfig}

\begin{document}

\begin{frontmatter}

\title{Projectile structure effects in the Coulomb breakup of 
one-neutron halo nuclei}

\author[Saha]{Rajdeep Chatterjee \thanksref{email1}},
\author[Saha]{Prabir Banerjee \thanksref{email2}} and
\author[Saha]{Radhey Shyam \thanksref{email3}}

\address[Saha]{Theory Group, Saha Institute of Nuclear Physics,
1/AF Bidhan Nagar, \\
Calcutta - 700 064, INDIA.}


\thanks[email1]{e-mail: raja@tnp.saha.ernet.in}
\thanks[email2]{e-mail: prabir@tnp.saha.ernet.in}
\thanks[email3]{e-mail: shyam@tnp.saha.ernet.in}

\begin{abstract}
We investigate the Coulomb breakup of neutron rich nuclei $^{11}$Be
and $^{19,17,15}$C within a theory developed in 
the framework of Distorted Wave Born Approximation. Finite range 
effects are included by a local momentum approximation, which allows 
incorporation of realistic wave functions for these nuclei in our 
calculations. Energy and angular as well as parallel  
momentum distributions of the fragments emitted in the breakup of these
nuclei on heavy targets have been calculated using 
several structure models for their ground state. Comparison
with the available experimental data shows that the results are selective
about the ground state wave function of the projectile. Our investigations
confirm that the nuclei $^{11}$Be, $^{19}$C and $^{15}$C
have a one-neutron halo structure in their ground states. However,
for $^{17}$C such a structure appears to be less likely.  Calculations
performed within our method have also been compared with those   
from an adiabatic model and the results are discussed. 
   
\end{abstract}


\end{frontmatter}

\renewcommand{\vec}[1]{\mbox{\boldmath$#1$\unboldmath}}
\newcommand{\cp}{\chi^{(+)}}
\newcommand{\cm}{\chi^{(-)*}}
\newcommand{\cmm}{\chi^{(-)}}
\newcommand{\vv}{V_{bc}({\bf r}_1)}
\newcommand{\ri}{{\bf r}_i}
\newcommand{\ro}{{\bf r}_1}
\newcommand{\ak}{{\bf k}_a}
\newcommand{\bq}{{\bf k}_b}
\newcommand{\lB}{\hat{l}\beta_{lm}}
\newcommand{\B}{\beta_{lm}}
\newcommand{\rc}{{\bf r}_c}
\newcommand{\cq}{{\bf k}_c}
\newcommand{\we}{\Psi^{(+)}_a(\xi_a,{\bf r}_1,{\bf r}_i)}
\newcommand{\bm}{\bibitem}
\newcommand{\fa}{ _2F_1(1-i\eta_a,1-i\eta_b;2;D(0))} 
\newcommand{\fB}{ _2F_1(-i\eta_a,-i\eta_b;1;D(0))}
 
\section{Introduction}

It has now been well established that close to the neutron drip line,
there exist nuclei having one, or some times two, very loosely bound
valence neutrons extending too far out in space with respect to a dense
charged core~\cite{tani85}. The properties of these neutron halo nuclei
\cite{han87} have been reviewed by several authors (see e.g.
\cite{mue93,rii94,han95,tani96}). The halo systems, which involve a new 
form of nuclear matter, are characterized by
large reaction and Coulomb dissociation cross sections
\cite{tani88,sack93,nak1,nak2}. Moreover, in the breakup reactions
induced by these nuclei, the   
angular distributions of neutrons measured in coincidence with the
core nuclei~\cite{ann90,mar96} are strongly forward peaked and  
the parallel momentum distributions of the core fragment have  
very narrow widths \cite{orr95,mex,baz,bm92,ps95,bau98}.
Due to their strikingly different properties as compared to those of
the stable nuclei, such systems provide a stringent test of the nuclear
structure models developed for the latter.

The Coulomb dissociation is a significant reaction
channel in the scattering of halo nuclei from a stable heavy target
nucleus. It provides a convenient tool to investigate their structure.
For instance, it would place constraints on their electric dipole response
\cite{nak1,nak2,ber88,ber91}. Of course, in the Serber \cite{serb47} 
type of models \cite{neg96,han96}, 
the breakup cross sections are directly related to the momentum space
wave function of the projectile ground state. The studies of the Coulomb
dissociation of weakly bound nuclei are also of interest due to their
application in determining the cross sections of the astrophysically
interesting radiative capture reactions at solar temperatures
\cite{baur94}.

The Coulomb dissociation of halo nuclei has been investigated by several
authors recently, using a number of different theoretical 
approaches. A semiclassical coupled channel formalism has been used 
by authors of Ref. \cite{cant93}, while in Refs. \cite{kid94,ber93} 
the relative motion of the core and the 
valence particle is described by a time dependent Schr\"odinger equation.
The results within these approaches depend on the range of the impact
parameter associated with the straight line trajectories used to describe 
the motion of the projectile in the field of target nuclei. However,
in these studies the emphasis was on investigating the dynamics
of the Coulomb dissociation, and not the structure of the projectile ground
state which was assumed to have some very simple zero range (ZR) form.
Similar assumption for the projectile structure was also made in other
semiclassical~\cite{han96,ann} and prior form distorted wave Born
approximation (DWBA) calculations~\cite{bert91}. 

On the other hand, the post form DWBA theory of the breakup reactions
incorporates the details of the ground state structure of the projectile 
in the breakup amplitude~\cite{shyam84}. However, in an earlier application
\cite{shyam92} of this theory to calculate the breakup of the halo nuclei,
the simplifying approximation of the zero range interaction was used.
This approximation necessarily excludes the use of this theory
to describe the breakup of non-$s$ -- wave projectiles. 
Moreover its applicability is
questionable even for $s$ -- wave projectiles at higher beam energies
\cite{shyam85}. Therefore, to investigate the details of the projectile
structure through breakup reactions within this theory, the inclusion
of the finite range effects is necessary. It may be noted that in a
recently formulated adiabatic model~\cite{toste98} of the Coulomb breakup
reactions, where it is assumed that the projectile excitation is
predominantly to the states of the low internal energy, the details 
of the ground state wave function also enters in the transition 
amplitude.
   
In this paper, we present a theoretical model to describe the Coulomb breakup
of a projectile within the framework of the post form 
DWBA where finite range effects are included, approximately, via a
local momentum approximation (LMA)~\cite{braun74a,braun74b}.
The exact treatment  
of the finite range effects within this theory, although desirable, is
much too complicated as it would lead to six dimensional integrals
involving functions which are oscillatory asymptotically. The LMA 
leads to two simplifying features. First, it factorises the dynamics of the 
reaction from the structure effects of the projectile and second, it
results in an amplitude where the term describing the dynamics of the 
process is the same as that evaluated in the ZR approximation. We present
the application of this theory to the Coulomb breakup of neutron rich nuclei $^{11}$Be and  $^{15,17,19}$C on a number of
heavy targets. We attempt to
put constraints on the ground state structure of these nuclei by
analyzing almost all the available data on the energy, angular and
longitudinal momentum distributions of fragments using 
various configurations of the projectile ground state.  

Our formalism is presented in section {\bf 2}. In
section {\bf 3} we present and discuss the results of our calculations 
for various observable for the reactions mentioned above.  
The summary and the conclusions of our work
are described in section {\bf 4}. The validity of the 
approximate method used by us to include the finite range effects is
presented in appendix A.

\section{Formalism}
   
We consider the reaction $ a + t \rightarrow b + c + t $, where the 
projectile $a$ breaks up into fragments $b$ (charged) 
and $c$ (uncharged) in the Coulomb
field of a target $t$. The coordinate system chosen is shown in Fig. 1.
\begin{figure}[ht]
\begin{center}
\mbox{\epsfig{file=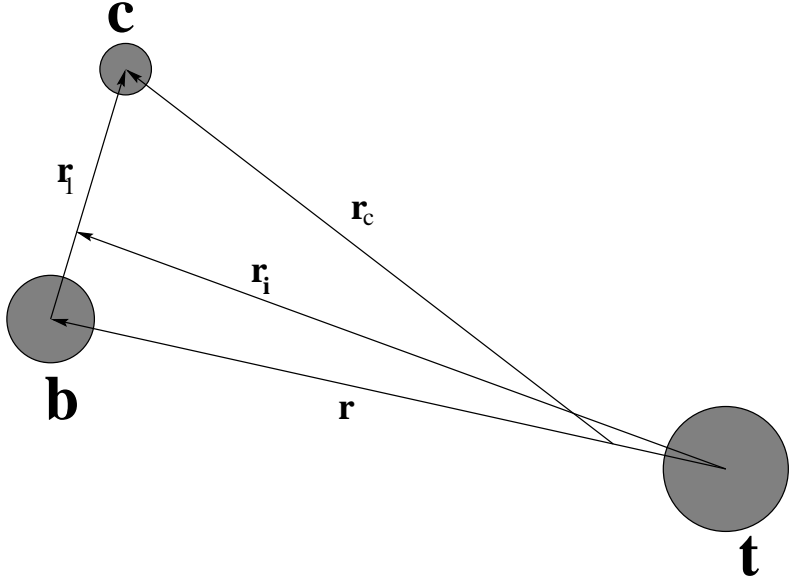,width=.75\textwidth}}
\end{center}
\caption{The three-body coordinate system. $b$, $c$ and $t$ represent the
charged core, valence neutron and target respectively. } 
\label{fig:figa)}
\end{figure}

The position vectors satisfy the following relations
\begin{eqnarray}
{\bf r} &=& \ri - \alpha\ro,~~ \alpha = {m_c\over {m_c+m_b}}   \\
\rc &=& \gamma\ro +\delta\ri, ~~ \delta = {m_t\over {m_b+m_t}}, 
~~  \gamma = (1 - \alpha\delta)    
\end{eqnarray}
The exact post form $T$ - matrix for this case is
\begin{eqnarray}
T & = &\langle
\chi^{(-)}_b({\bq},{\bf r})\Phi_b(\xi_b)\chi^{(-)}_c({\cq},{\rc})\Phi_c(\xi_c)
|\vv|\we \rangle,
\end{eqnarray} 
 where $\chi's$ are the distorted waves for relative motions of 
$b$ and $c$ with respect to $t$ and the center of mass (c.m.) of $b+t$
system respectively, and $\Phi's$
are the internal state wave functions of the concerned particles with 
internal coordinates $\xi$. $\we$ is
the exact three-body scattering wave function of the projectile  
with a wave vector $\ak$, with outgoing wave boundary condition.  
$\bq$, $\cq$ are Jacobi wave vectors of $b$
and $c$ respectively in the final channel of the reaction. $\vv$ is the 
interaction between $b$ and $c$. The charged fragment $b$ interacts with
the target by a point Coulomb interaction and hence 
$\chi^{(-)}_b({\bq},{\bf r})$ is a Coulomb distorted wave with incoming
wave boundary
condition. For pure Coulomb breakup, the interaction between
the target and uncharged fragment $c$ is zero and hence
 $\chi^{(-)}_c({\cq},{\rc})$ is
replaced by a plane wave.  
 
In the distorted wave Born approximation (DWBA), we write 
\begin{eqnarray}
\we = \Phi_a(\xi_a,\ro)\cp_a(\ak,\ri),
\end{eqnarray}
where $\Phi_a(\xi_a,\ro)$ represents the bound state wave function of the 
projectile having its radial and angular parts as $u_{\ell}(r_1)$
and ${Y_{\ell m}({\hat{\bf r}}_1)}$
respectively, which are associated with the
relative motion of $b$ and $c$.  
$\cp_a(\ak,\ri)$ is the Coulomb distorted 
scattering wave describing the relative motion of the c.m.
of the projectile with respect to the target with outgoing wave boundary
condition. The assumption inherent in Eq. (4) is that the breakup channels 
are very weakly coupled and hence this coupling needs to be treated only in the 
first order. The transition amplitude (written in the integral form) is
now given by 
\begin{eqnarray}
 T  = \int\int\int && d\xi d\ro d\ri \cm_b(\bq,{\bf r})\Phi^*_b(\xi_b)  
 e^{-i\cq.\rc}\Phi^*_c(\xi_c)\vv  \nonumber  \\
 &\times& {}\Phi_a(\xi_a,\ro)\cp_a(\ak,\ri).   
\end{eqnarray}
The integrals over the internal coordinates $\xi$ give
\begin{eqnarray}
 &&\int d\xi\Phi^*_b(\xi_b)\Phi^*_c(\xi_c)\Phi_a(\xi_a,\ro)  \nonumber \\
 &&= \sum_{\ell mj\mu} \langle \ell mj_c\mu_c|j\mu\rangle 
 \langle j_b\mu_bj\mu|j_a\mu_a\rangle i^\ell u_\ell(r_1)
 {Y_{\ell m}({\hat{\bf r}}_1)}.
\end{eqnarray}
In Eq. (6) $\ell$ is the relative motion angular momentum
between $b$ and $c$ . This
is coupled to spin of $c$ and the resultant $j$ is coupled to spin of
(the inert core) $b$ to get the spin of $a$ ($j_a$).
Using Eq. (6), the $T$-matrix can be written as  
\begin{eqnarray} 
 T & = & \sum_{\ell mj\mu} \langle \ell mj_c\mu_c|j\mu\rangle 
 \langle j_b\mu_bj\mu|j_a\mu_a\rangle i^\ell
 \hat{\ell}\beta_{\ell m}(\bq,\cq;\ak),
\end{eqnarray}
where
\begin{eqnarray}
\hat{\ell}\beta_{\ell m}(\bq,\cq;\ak)  =  
\int\int && d\ro d\ri\cm_b(\bq,{\bf r})e^{-i\cq.\rc}
 \vv \nonumber \\
 && \times {} \phi^{\ell m}_a(\ro) \cp_a(\ak,\ri),    
\end{eqnarray}
with $\beta_{\ell m}$ being the reduced $T$ -- matrix and
 ${\hat \ell} = \sqrt{2\ell + 1}$. We have 
written $\phi^{\ell m}_a(\ro) = u_\ell(r_1){Y_{\ell m}({\hat{\bf r}}_1)}$. 

It may be noted that the reduced amplitude $\beta_{\ell m}$ involves
a six dimensional integral which makes its evaluation quite complicated.
The problem gets further acute due to the fact that the integrand involves
three scattering waves which have oscillatory behavior asymptotically.
Therefore, several approximate methods have been used in the literature
to avoid the evaluation of six dimensional integrals.
 In the zero range approximation
(ZRA) \cite{satchler} one assumes 
\begin{eqnarray}
 \vv\phi^{\ell m}_a(\ro) & = & D_{0}\delta(\ro),
\end{eqnarray}
 where $D_0$ is the usual zero range constant. This approximation 
reduces the six dimensional integral in Eq.(8) to a three-dimensional 
one. The corresponding amplitude is written as 
\begin{eqnarray}
\beta^{ZR}_{00} & = & D_{0}
\langle \chi^{(-)}_b(\bq,\ri) e^{i\delta\cq.\ri}|\cp_a(\ak,\ri)\rangle.
\end{eqnarray}
In Eq. (10), the details of the projectile structure
enter in the reaction amplitude only as a multiplicative constant $D_0$. 
However, ZRA necessarily restricts the relative
motion between $b$ and $c$ in the projectile $a$ to $s$ -- state only.
Even for such cases, this approximation may not be satisfied for
heavier projectiles and at higher beam energies \cite{shyam85}.
 
Baur and Trautmann (BT) \cite{BT} have proposed an alternative approximate
scheme in which the projectile c.m. coordinate in the corresponding 
distorted wave in Eq. (8) is replaced by that of the core-target system, 
i.e. $\ri \approx {\bf r}$. With this approximation the amplitude
$\beta_{\ell m}$ splits into two factors each involving a three dimensional
integral 
\begin{eqnarray}
{\hat \ell}\beta^{BT}_{\ell m} & = & {\langle e^{i\cq.\ro}|V_{bc}|
\phi^{\ell m}_a(\ro)\rangle}
\langle \chi^{(-)}_b(\bq,{\bf r}) e^{i\delta\cq.{\bf r}}|\cp_a(\ak,{\bf r})
\rangle.
\end{eqnarray}
The first term, known as the form factor, depends on the structure of
the projectile through its ground state wave function
 $\phi^{\ell m}_a({\bf r_1})$. 
The second term involves the dynamics of the reaction.
This amplitude (which will be referred to as the BT amplitude),
used originally to study the deuteron breakup at 
sub-Coulomb energies \cite{BT}, was applied to the calculations of
the Coulomb breakup of halo nuclei in \cite{shyam92}. This approximation,
which allows the  
application of the theory to non-$s$ -- wave projectiles,   
may seem to be justified if the c.m of the $b + c$ system
is shifted towards $b$ (which is indeed the case if $m_b \gg m_c$). 
However, the neglected piece of ${\bf r}_i$ ($\alpha \ro$) occurs in   
association with the wave vector $\ak$, whose magnitude
may not be negligible for the reactions at the higher beam energies. 
Therefore, contribution coming to the amplitude from the neglected part of
${\bf r}_i$ may still be substantial. 
     
An approximate way of taking into account the finite range effects in
the post form DWBA theory is provided by the local momentum approximation
\cite{shyam85,braun74a}. The attractive feature of this approximation
is that it leads to the factorization of the amplitude $\beta_{\ell m}$
similar to that obtained in the BT case.  We use this
approximation to write the Coulomb distorted wave of particle $b$ in the 
final channel as 
\begin{eqnarray}
\chi^{(-)}_b(\bq,{\bf r}) & = & e^{-i\alpha{\bf K}.\ro}
                           \chi^{(-)}_b(\bq,\ri). 
\end{eqnarray}
Eq. (12) represents an exact Taylor series expansion about ${\bf r}_i$ if 
${ {\bf K}}( = -i\nabla_{{\bf r}_i})$ is treated exactly. However,
this is not done in the LMA scheme. Instead, the
magnitude of the local momentum here is taken to be 
\begin{eqnarray} 
{ {K}}(R) = {\sqrt {{2m\over \hbar^2}(E - V(R))}},
\end{eqnarray}
where $m$ is the reduced mass of the $b-t$ system,
$E$ is the energy of particle $b$ relative to the target in the
c.m. system and $V(R)$ is the  Coulomb potential
between $b$ and the target at a distance $R$. Therefore, 
the local momentum ${{\bf K}}$ is evaluated at some distance
$R$, and its magnitude is held fixed for all the values of ${\bf r}$. 
As is discussed in appendix A, the results of our calculations
are almost independent of the choice of the direction the local momentum.
Therefore, we have taken the directions of ${{\bf K}}$ and
${\bf k_b}$ to be the same in all the calculations presented in this
paper. It may be noted that in the calculations presented in  Ref. \cite{pb1},
the LMA was applied to the Coulomb distorted wave of the projectile 
channel, where making a choice of the direction of the local momentum is
some what  complicated due to the fact
that the directions of both the fragments in the final channel has to be
brought into the consideration. Detailed discussion on the validity of the
local momentum approximation is presented in appendix A. 

Substituting Eq. (12) into Eq. (8) we get the following factorized form of
the amplitude  
\begin{eqnarray}
{\hat \ell}\beta^{FRDWBA} _{\ell m}& = &
\langle e^{i(\gamma\cq - \alpha {\bf K}).\ro}|V_{bc}|
\phi^{\ell m}_a(\ro)\rangle \nonumber \\
        & \times &\langle \chi^{(-)}_b(\bq,\ri) e^{i\delta\cq.\ri}|
\cp_a(\ak,\ri)\rangle.
\end{eqnarray}
Eq. (14) (which will be referred to as the FRDWBA amplitude in the following) 
looks like Eq (11) of the BT theory but with the
very important difference that the  
form factor is now evaluated at the momentum transfer  
($\gamma {\bf k}_c - \alpha{ {\bf K}}$), and not at the valence particle 
momentum ${\bf k_c}$. The two momenta could be quite different for 
cases of interest in this paper.  
The second term, involving the dynamics part of the reaction, is the same
in both the cases. Therefore, the breakup amplitude obtained in 
BT approximation differs from that of FRDWBA by a factor 
\begin{eqnarray}
F_r  = {{\beta^{BT}_{\ell m}}\over {\beta^{FRDWBA}_{\ell m}}} = 
{{\langle e^{i\cq.\ro}|V_{bc}|\phi^{\ell m}_a(\ro)\rangle}\over
{\langle e^{i(\beta\cq - \alpha {\bf K}).\ro}|V_{bc}|\phi^{\ell m}_a(\ro)\rangle}}
\end{eqnarray}

Recently, a theory of the Coulomb breakup has been developed within an 
adiabatic (AD) model \cite{toste98,jal97}, where one assumes that the 
excitation of the projectile is such that the relative energy ($E_{bc}$)
of the $b-c$ system
is much smaller than the total incident energy, which allows 
$E_{bc}$ to be replaced by the constant separation energy of the
fragments in the projectile ground state. It was shown in \cite{jal97}
that under these conditions the  wave function $\we$
has an exact solution as given below
\begin{eqnarray}
\Psi_a^{(+),AD}(\xi_a,{\bf r}_1,{\bf r}_i) & = &
 \Phi_a(\xi_a,\ro)e^{i\alpha\ak.\ro}\cp_a(\ak,{\bf r})
\end{eqnarray}

Substituting $\Psi_a^{(+),AD}$ for $\Psi_a^{(+)}$ in Eq. (3)
leads to the reduced amplitude:
\begin{eqnarray}
{\hat \ell}\beta^{AD}_{\ell m} & = &
\langle e^{i(\cq - \alpha\ak).\ro}|V_{bc}|\phi^{\ell m}_a(\ro)\rangle
\langle \chi^{(-)}_b(\bq,{\bf r}) e^{i\delta\cq.{\bf r}}|\cp_a(\ak,{\bf r})
\rangle.
\end{eqnarray}
This amplitude differs from those of BT as well as  FRDWBA only in
the form factor part (first term), which is evaluated here at the momentum 
transfer $(\cq - \alpha \ak)$.  
Eq. (17) can also be obtained in the DWBA model 
by making a local momentum
approximation to the Coulomb distorted wave in 
the initial channel of the reaction, and by evaluating the local momentum
at $R = \infty$ with its direction being the
same as that of the projectile \cite{pb1}. The adiabatic model does
not make the weak coupling approximation of the DWBA. However,
it necessarily requires one of the
fragments (in this case $c$) to be chargeless. In contrast, 
the DWBA with the LMA could, in principle, be applied to cases where both
the fragments $b$ and $c$ are charged \cite{shyam85} (of course the dynamical
part of the amplitude can not be expressed in terms of the simple
bremsstrahlung integral in this case as discussed below).
While the effect of nuclear breakup in the adiabatic model description of elastic
scattering of the loosely bound projectile has been calculated in 
Ref. \cite{jal97,jc97,ron98}, the nuclear part of the amplitude for  breakup
reactions is yet to be calculated within this model. At the same time, 
the nuclear breakup cross section has not been calculated within the 
FRDWBA theory also (although it is straightforward to do so). These
calculations will be presented in a future publication.  

The triple differential cross section of the reaction is given by
\begin{eqnarray}
{{d^3\sigma}\over{dE_bd\Omega_bd\Omega_c}} & = & 
 {2\pi\over{\hbar v_a}}\rho(E_b,\Omega_b,\Omega_c)\sum_{\ell m}|\beta_{\ell m}|^2,
\end{eqnarray}
where $\rho(E_b,\Omega_b,\Omega_c)$ is the appropriate 
\cite{fuchs} three-body phase space factor, given by
\begin{eqnarray}
\rho(E_b,\Omega_b,\Omega_c) & = &{h^{-6}m_bm_cm_tp_bp_c \over 
m_t+m_c-m_c{{\bf k_c.(k_a-k_b)} \over k_c^2}}, 
\end{eqnarray}
with ${\bf k}_a,{\bf k}_b$ and ${\bf k}_c$ being evaluated in
the appropriate frame of reference. $v_a$
is the relative velocity of the projectile in the initial channel. $p$
in Eq. (19) are the linear momenta which are related to wave numbers $k$
by $p \,\, = \,\, \hbar k$.

Substituting the following expressions for the Coulomb distorted waves
\begin{eqnarray}
\cm_b(\bq,{\ri}) & = &
 e^{-\pi\eta_b/2}\Gamma(1 + i\eta_b) e^{-i\bq.\ri} 
 {_1F_1(-i\eta_b, 1, i(k_b r_i + \bq.\ri))},   \\
                                          \nonumber \\
\cp_a(\ak,\ri) & = &
 e^{-\pi\eta_a/2}\Gamma(1 + i\eta_a) e^{i\ak.\ri} 
 {_1F_1(-i\eta_a, 1, i(k_a r_i - \ak.\ri))}
\end{eqnarray}
in Eq. (14), one gets for the triple differential cross section
\begin{eqnarray}
{{d^3\sigma}\over{dE_bd\Omega_bd\Omega_c}} = {2\pi\over{{\hbar}v_a}}
\rho(E_b,\Omega_b,\Omega_c) 
{4\pi^2\eta_a\eta_b\over (e^{2\pi\eta_b}-1)(e^{2\pi\eta_a}-1)}|I|^2  
4\pi\sum_{\ell} |Z_{\ell}|^2.
\end{eqnarray}
In Eqs. (20 -- 22) $\eta$'s are the Coulomb parameters for the concerned
particles. In Eq. (22)
$I$ is the Bremsstrahlung integral \cite{bem} which can be evaluated in a 
closed form:  
\begin{eqnarray}
I &=& -i{\Big[}B(0){\Big(}{{dD}\over{dx}}{\Big)}_{x=0}(-\eta_a\eta_b)\fa  \nonumber \\
& + & {\Big(}{{dB}\over{dx}}{\Big)}_{x=0} {\fB} {\Big]}
\end{eqnarray}
where
\begin{eqnarray}
B(x) = {4\pi\over{k^{2(i\eta_a+i\eta_b+1)}}}
{\Big[}(k^2 - 2{\bf k}.\ak -2xk_a)^{i\eta_a}
(k^2 - 2{\bf k}.\bq -2xk_b)^{i\eta_b}{\Big]},
\end{eqnarray}
\begin{eqnarray}
D(x) = {2k^2(k_ak_b+\ak.\bq)-4({\bf k}.\ak+xk_a)({\bf k}.\bq+xk_b)\over
{(k^2 - 2{\bf k}.\ak -2xk_a)(k^2 - 2{\bf k}.\bq -2xk_b)}}
\end{eqnarray}
with ${\bf k} = \ak - \bq -\delta\cq$.
$Z_{\ell}$ contains the projectile structure information and is given by
\begin{eqnarray}
Z_{\ell} = \int dr_1 r^2_1 j_{\ell} (k_1 r_1)\vv u_{\ell} (r_1),
\end{eqnarray}
with $k_1 = |\gamma\cq - \alpha {\bf K}|$.

\section{Calculations and discussions}

Apart from the distance at which local momentum is calculated (which is
taken to be 10 $fm$ as discussed in appendix A) and its direction
(described already in the previous section), the only other input
to our calculations is the radial part of the projectile ground state
wave function. The forms for this as chosen by us for the projectiles
considered in this paper are discussed in the following.
  
\subsection{Calculations for reactions induced by $^{11}$Be}

For the ground state of $^{11}$Be, we have considered the following 
configurations : (a) a $s$ -- wave valence neutron coupled to $0^+$  
$^{10}$Be core ($^{10}$Be$(0^+) \otimes 1s_{1/2}\nu$) with a   
one-neutron separation energy ($S_{n-^{10}Be}$) of 504 keV and a spectroscopic
factor (SF) of 0.74 \cite{jon,auton70,zwieg79},
(b) a $d$ -- wave valence neutron coupled to $2^+$ $^{10}$Be core
($^{10}$Be$(2^+)\otimes0d_{5/2}\nu$) with a binding energy of 3.872 MeV 
(which is the sum of energy of the excited 2$^+$ core (3.368 MeV) 
and $S_{n-^{10}Be}$) and (c) an admixture of these two configurations with
spectroscopic factors of 0.74 and 0.17 respectively
\cite{aumann00}. In each case, the single particle wave function,  
is constructed by assuming the valence neutron-$^{10}$Be 
interaction to be of Woods-Saxon type whose depth is adjusted to reproduce
the corresponding value of the binding energy with fixed values of the radius
and diffuseness parameters (taken to be  1.15 $fm$ and 0.5 $fm$
respectively). For configuration (a), this
gives a potential depth of 71.03 MeV, a root mean square (rms) radius 
for the valence neutron of 6.7 $fm$ and a rms radius for $^{11}$Be
of 2.91 $fm$ when the size of the $^{10}$Be core is taken to
be 2.28 $fm$ \cite{jal96}. In some cases we have also used a wave function
for the $s$ -- state of $^{11}$Be calculated within a dynamical core
polarization (DCP) model
\cite{len99}.

\subsubsection{Energy distributions of the fragments}
\begin{figure}[ht]
\begin{center}
\mbox{\epsfig{file=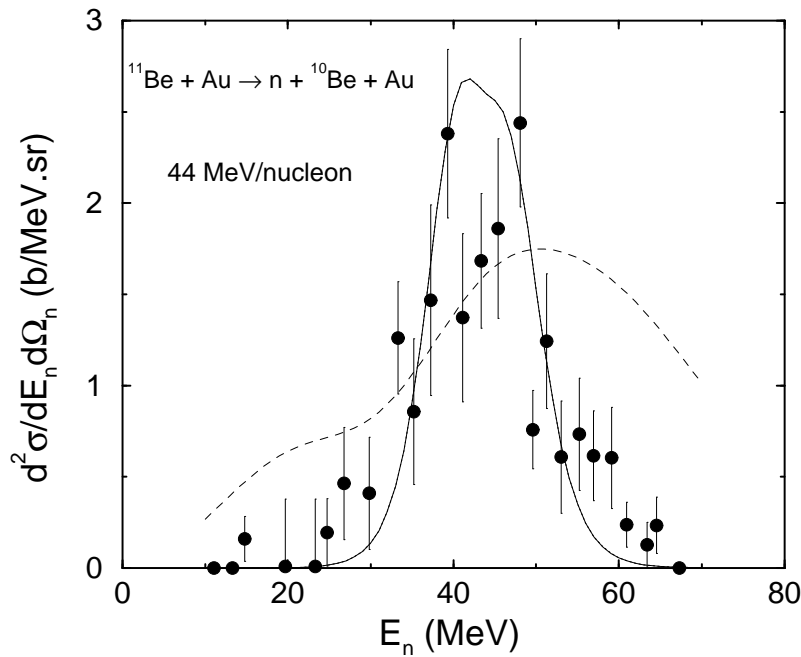,width=.75\textwidth}}
\end{center}
\caption{Neutron energy distributions for the breakup of $^{11}$Be on the 
Au target at the beam energy of 
44 MeV/nucleon calculated within the FRDWBA using configurations
(a) (dotted line), (b) (dashed line, plotted after multiplying the
actual results by a factor of 1000) and (c) solid line for the ground state
of $^{11}$Be. The results obtained with configurations (a) and (c) are 
indistinguishable from each other. The experimental data are  
taken from \protect\cite{ann}.} 
\label{fig:figb)}
\end{figure}

In Fig. 2, we present the results of our FRDWBA calculations for the 
double differential cross section ($d^2\sigma/dE_n d\Omega_n$)
as a function of the neutron energy at the
neutron angle of $1^\circ$, in the
breakup of $^{11}$Be on a gold target at the beam energy of
44 MeV/nucleon. The experimental data are taken from \cite{ann}.
The results obtained with configurations (a), (b) and (c)
for the ground state of $^{11}$Be are shown by dotted, dashed and full
lines respectively. The results of configuration (b) are plotted after
multiplying the actual numbers by a factor of 1000. The cross sections
obtained with configurations (a) and (c) are indistinguishable
from each other. Thus these two configurations produce almost identical
results for the Coulomb dissociation of $^{11}$Be. 
In the following, we have used 
configuration (a) for the ground state state of $^{11}$Be in all the
calculations. 

\begin{figure}[ht]
\begin{center}
\mbox{\epsfig{file=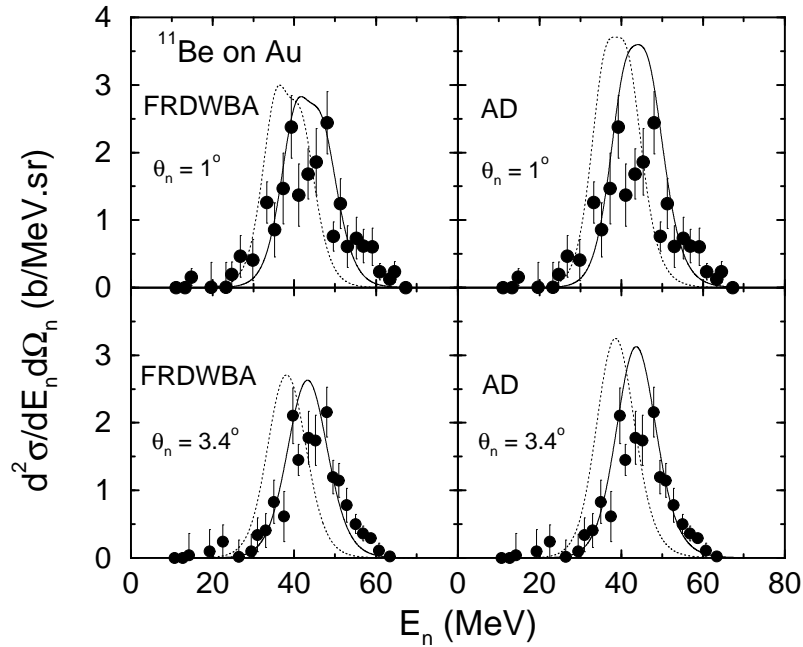,width=.75\textwidth}}
\end{center}
\caption{Neutron energy distributions for the breakup of $^{11}$Be on Au at
beam energies of 37 MeV/nucleon (dotted lines) and 44 MeV/nucleon (solid 
lines), calculated using configuration (a) with single particle wave
functions within the FRDWBA 
and the AD models. The top half of the
figure is for $\theta_n = 1^{\circ}$, while the bottom half is for 
$\theta_n = 3.4^{\circ}$. The experimental data are taken from 
\protect\cite{ann}.} 
\label{fig:figc}
\end{figure}

In Fig. 3, we present a comparison of our calculation with the data 
(taken from \cite{ann}) for the energy distribution of the neutron emitted
in the same reaction as in Fig. 2 at two forward angles.
Calculations performed within both FRDWBA and AD model
are shown in this figure. The same configuration  
for the $^{11}$Be ground state has been used in both the cases. 
The beam energy in this experiment \cite{ann} varies between 36.9 -- 44.1 
MeV/nucleon.  To take into account this spread,
 we have performed calculations at both
its upper (44 MeV/nucleon) (solid line) and lower ends
(37 MeV/nucleon) (dotted line). Although these data have large statistical
errors, the calculations performed at 44 MeV/nucleon are in better agreement
with the experimental values.
Thus our calculations may serve to remove the uncertainty
in the data in this regard. It should also be noted that the AD model 
calculations over-predict the experimental cross sections in the peak region.
\begin{figure}[ht]
\begin{center}
\mbox{\epsfig{file=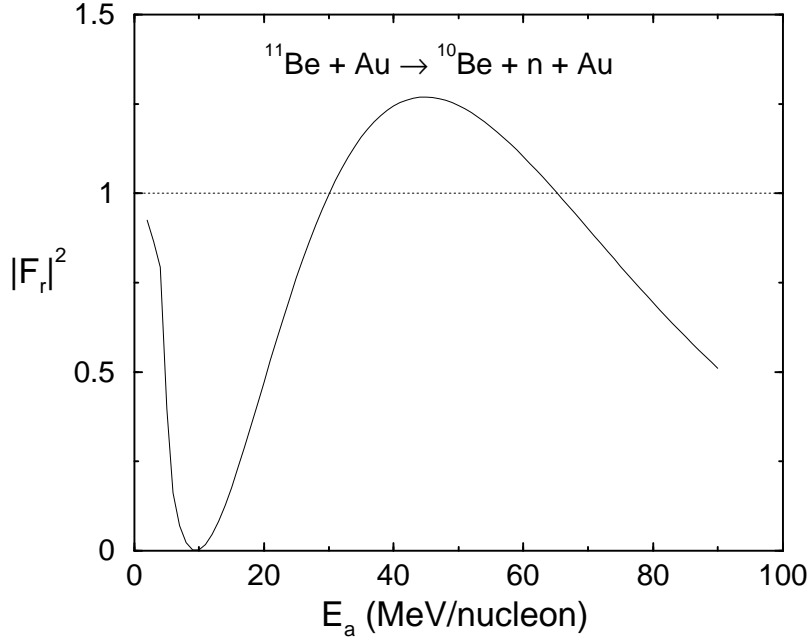,width=.75\textwidth}}
\end{center}
\caption{Modulus square of the ratio $F_r$ defined by Eq. (15) as a 
function of the beam energy for the same reaction as in Fig. 2
corresponding to forward emission angles of the breakup  fragments as 
discussed in the text.}
\label{fig:figd}
\end{figure}

As discussed earlier, the cross sections obtained in the BT theory
differs from that of the FRDWBA by the modulus square of the
factor $F_r$ as defined by Eq. (15). In fig. 4, we have shown the
beam energy dependence of $|F_r|^2$ for the same reaction as in Fig. 2,
for a set of forward angles of the outgoing fragments 
($\theta _b = 1^{\circ}$, $\theta _c = 1^{\circ}$ and 
$\phi _c = 1^{\circ}$). We can see that this quantity is close to unity only 
at the sub-Coulomb beam energies (of course at higher beam
energies it crosses twice the line representing the value 1).
Therefore, the BT and FRDWBA calculations are expected to produce
similar results at very low incident energies. Depending upon
the beam energy, the BT results can be larger or smaller than 
those of the FRDWBA.
\begin{figure}[ht]
\begin{center}
\mbox{\epsfig{file=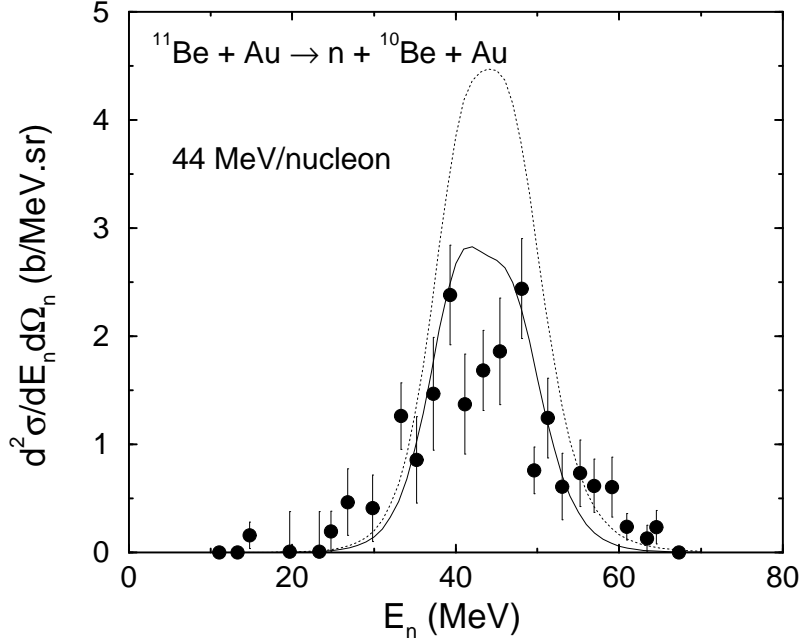,width=.75\textwidth}}
\end{center}
\caption{
The energy distribution of the neutrons observed in the breakup of
$^{11}$Be on the Au target at the beam energy of 44 MeV/nucleon at the
neutron angle of 1$^\circ$, calculated within the FRDWBA (solid line) and 
BT (dashed line) theories.}
\label{fig:fige}
\end{figure}
  
A comparison of the results of the FRDWBA (solid line) and the BT 
(dotted line) calculations for the same reaction as in Fig. 2 
is presented in Fig. 5. The bombarding energy in this case is 
44 MeV/nucleon. We can see that the BT results are larger
than those of the FRDWBA in almost entire region of neutron energies. This
is to be expected from the results shown in Fig. 4, where the quantity
$|F_r|^2$ is larger than unity at this beam energy. It may be noted that
the FRDWBA results provide a reasonable good description of the data,
particularly in the peak region while BT calculations overestimate
them. 

\begin{figure}[ht]
\begin{center}
\mbox{\epsfig{file=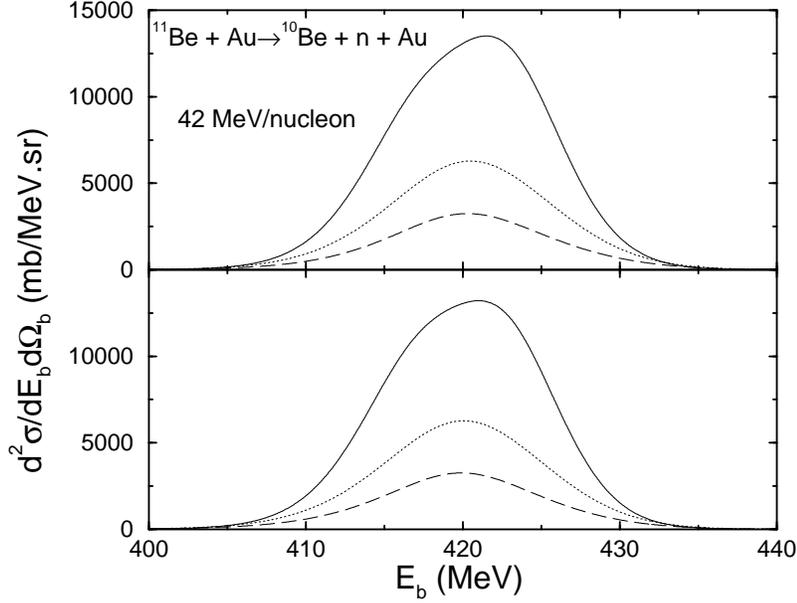,width=.75\textwidth}}
\end{center}
\caption{
Energy distribution of the $^{10}$Be fragment in the breakup of
$^{11}$Be on Au at 42 MeV/nucleon beam energy corresponding to the
$^{10}$Be angles of 1$^\circ$ (solid line), 2$^\circ$ (dotted line) and
3$^\circ$ (dashed line). The FRDWBA and adiabatic model results are shown 
in the upper and lower parts of the figure respectively.
}
\label{fig:figf}
\end{figure}
 
In Fig. 6, we show the energy distributions of the charged fragment
$^{10}$Be, calculated  within the FRDWBA (upper part) and the AD model
(lower part) for the breakup of $^{11}$Be on the Au target at the
beam energy of 42 MeV/nucleon at three angles of
1$^{\circ}$, 2$^{\circ}$ and 3$^{\circ}$. A noteworthy feature of
these results is that the peak position in both the calculations is
very close to the energy corresponding to the beam velocity. This
suggests that the charged fragment gets almost no post acceleration
in the final channel. This feature of the breakup of halo nuclei (which
is in contrast to the case of the breakup of stable isotopes \cite{shyam84}),
was first noted in Ref. \cite{shyam92}, and was corroborated later on
by the authors of Refs. \cite{cant93,kid94,ann,pb2}.
The reason for not observing the post acceleration effects, as put
forward by authors of Ref. \cite{shyam92}, is that due to their very
small binding energies the halo nuclei break up far away from the
distance of closest approach, which reduces greatly the effect of Coulomb
repulsion on the charged outgoing fragment. This argument has
received support from a recent calculation of the  Coulomb dissociation
potential of the deuteron which is shown  \cite{anders99} to have
a considerably large value even outside the charge density of the
target nucleus. A separate reason for not observing this effect
has been put forward by the authors of Ref.~\cite{ann}, according to which
these effects are small as the collision time is much less than the 
characteristic time for the disintegration of the halo. 

However, these quasiclassical arguments have been questioned by the authors
of Ref. \cite{ber93,esbe95}. Esbensen, Bertsch and Bertulani \cite{esbe95},
who include the higher order processes in the Coulomb dissociation of
halo nuclei $^{11}$Li and  $^{11}$Be by solving the three-dimensional
time-dependent Schr\"odinger equation, and find a magnitude
for the post acceleration effect which is quite appreciable for the $^{9}$Li
fragment in the breakup of $^{11}$Li and relatively somewhat smaller for the 
$^{10}$Be fragment in the breakup of $^{11}$Be. In the semiclassical
calculations of Baur, Bertulani and Kalassa \cite{bert92}, where breakup
is assumed to take place at some classical distance,
it is predicted that post acceleration effects should manifest itself
in the increase of the 
average momentum of the charged fragment
with the scattering angle. Indeed, the earlier measurements
of Nakamura {\em et al.} \cite{nak1} is consistent with this observation.

\begin{figure}[ht]
\begin{center}
\mbox{\epsfig{file=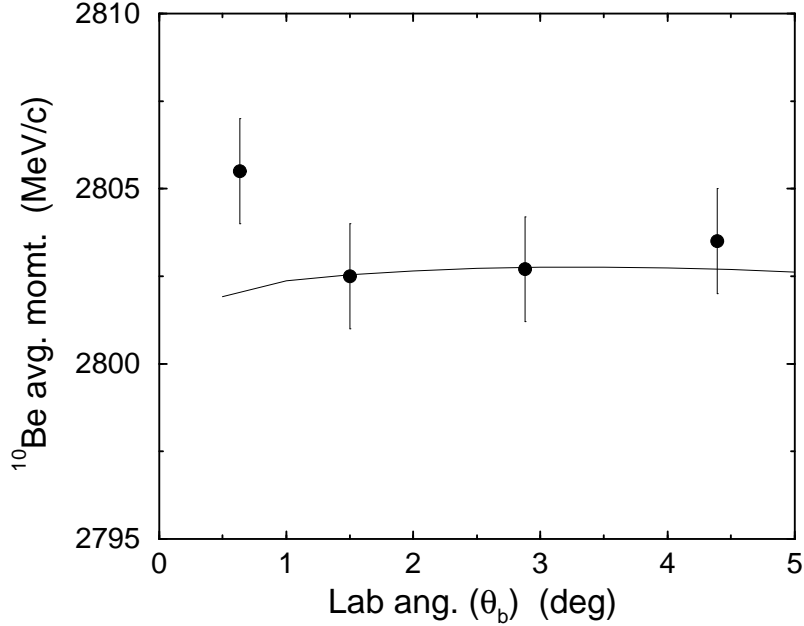,width=.75\textwidth}}
\end{center}
\caption{
Average momenta of the $^{10}$Be fragment in the breakup of $^{11}$Be
on Au at 42 MeV/nucleon beam energy as a function of its detection
angle. The data are taken from \protect\cite{nopa}.} 
\label{fig:figg}
\end{figure}

In a recent experiment \cite{nopa}, the momentum of the $^{10}$Be core
from the breakup of $^{11}$Be has been measured with sufficient precision
to verify the previously reported post acceleration effect. We have made a
comparison of our FRDWBA calculations with this data in  Fig. 7, where
 we show the calculated and experimental $^{10}$Be average
momenta (defined as $\sum p_b \frac{d^2\sigma}{dp_bd{\Omega_b}}/
\sum \frac{d^2\sigma}{dp_bd{\Omega_b}}$) as
a function of laboratory angle. 
In the semiclassical picture of Ref. \cite{bert92}, the impact
parameter decreases with the increase of the scattering angle, 
thereby making the Coulomb repulsion effects on the charged fragment stronger. 
Therefore, the post acceleration should show up in the increase of this average
momentum with the increase of the scattering angle. However, in both the 
experimental data of \cite{nopa} as well as our calculations no 
such increase is observed. Thus, neither the data of Ref. \cite{nopa}
nor our calculations support the results found in Ref. \cite{nak1}.
Therefore, the semiclassical picture presented in \cite{bert92} should
be viewed with caution. 
 
\subsubsection{Neutron Angular Distribution}
\begin{figure}[ht]
\begin{center}
\mbox{\epsfig{file=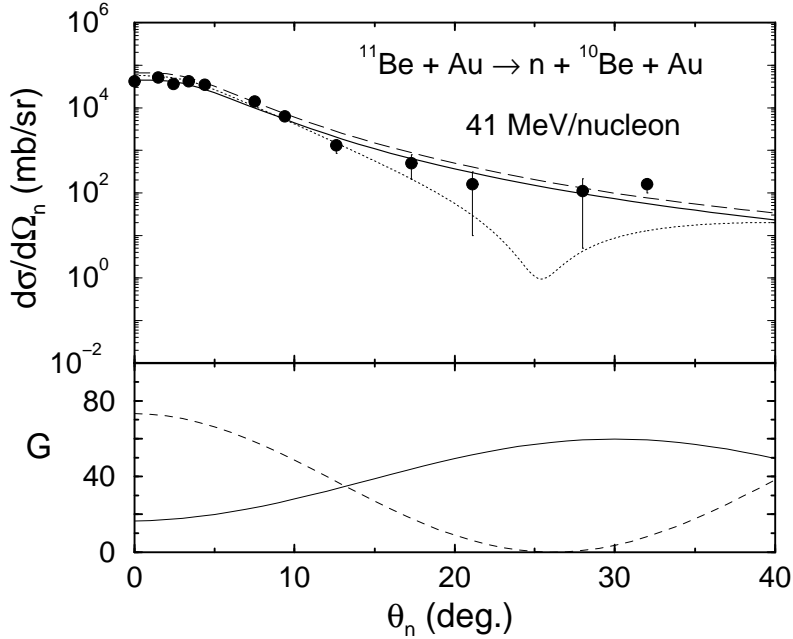,width=.75\textwidth}}
\end{center}
\caption{The calculated neutron angular distributions for 
the breakup of $^{11}$Be on a Au target at 41 MeV/nucleon beam energy
(upper half). The solid and the dotted
lines show the results of the calculation performed with the FRDWBA
and AD model with configuration (a) using single particle wave functions,
while the dashed line corresponds to the FRDWBA calculation done with the
DCP wave function for the ground state of $^{11}$Be as discussed in the text. 
The data are taken from \protect\cite{ann}. In the lower half, the quantity 
$G$ (see text) has been plotted against $\theta_n$. The solid line and 
the dashed lines show the results for the FRDWBA and the AD cases
respectively.} 
\label{fig:figh}
\end{figure}

The measured 
neutron angular distribution in the exclusive
$^{11}$Be + $A$ $\rightarrow$ $^{10}$Be + n +$A$ reaction below the
grazing angle is very narrow and is shown to be \cite{pb1,pb93}
dominated by the Coulomb breakup
process. This reflects the narrow width of the transverse momentum 
distribution of the valence neutron in the ground state of $^{11}$Be,
which is consistent with the presence of a neutron halo structure
in $^{11}$Be. In the top half of Fig. 8, we compare the calculated
and measured
exclusive neutron angular distribution $d\sigma/d\Omega_n$ as a
function of the neutron angle $\theta_n$ for the 
above reaction on a Au target at the beam energy of 41 MeV/nucleon. 
Calculations (where integrations over the core fragment energy is 
done in the range of 390 to 430 MeV, which contributes most to the 
cross section) performed within the FRDWBA (solid line) and the AD model
(dotted line) using the same  configuration for the $^{11}$Be ground state
are shown in this figure. Also shown (dashed line) here is the FRDWBA
calculation performed with the $^{11}$Be ground state wave function
obtained in the DCP model. We note that while the 
FRDWBA and the AD model results
agree with each other well below 12$^\circ$, the difference between the
two models starts becoming prominent as the angle increases beyond this
value, with the latter developing a dip around 25$^\circ$.
At small neutron angles, the FRDWBA calculation done with the
DCP wave function overestimates somewhat the measured neutron
angular distribution.

To understand the origin of the dip in the AD model calculations of  
the neutron angular distribution, we have
plotted the quantity $G (= \int d\theta_b\sin \theta _b|Z_{\ell}|^2 $) 
as a function of $\theta _n$ in the lower half of Fig. 8 
for both the FRDWBA (solid line) and the AD (dashed line) 
cases. The energy of fragment $b$ corresponds to the beam velocity. 
We can see that in the AD case $G$ becomes very small around 25$^{\circ}$,
which corresponds to a node in its form factor at the momentum
transfer related to this angle.
   
\subsubsection{Relative energy spectrum of fragments} 
\begin{figure}[ht]
\begin{center}
\mbox{\epsfig{file=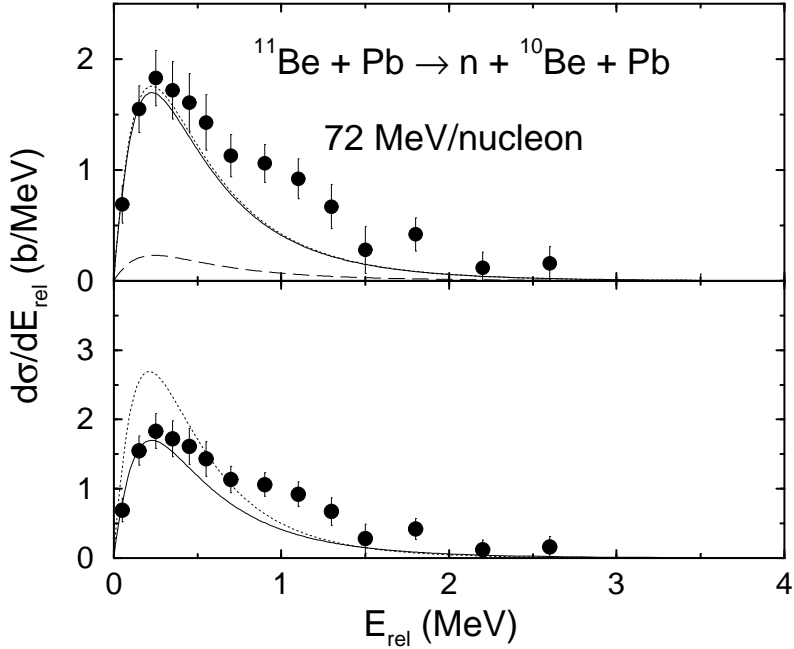,width=.75\textwidth}}
\end{center}
\caption{Relative energy spectra for the Coulomb breakup of $^{11}$Be on a
Pb target at 72 MeV/nucleon beam energy. The top half of the figure shows
the spectra obtained with a single particle wave function,
using the FRDWBA (solid line), the AD model (dotted line) and the 
BT approximation (dashed line). The bottom half shows the results of
FRDWBA calculations using single particle (solid line) and DCP
(dotted line) wave functions. The data are taken from \protect\cite{nak1}.} 
\label{fig:figi}
\end{figure}

The relative energy spectrum for the breakup of $^{11}$Be on a Pb target
at 72 MeV/nucleon is shown in Fig. 9.
The top half shows the results obtained with the FRDWBA (solid line) and 
the AD model (dotted line) using configuration (a) with the single
particle wave function for the $^{11}$Be
ground state, while the bottom half depicts the FRDWBA 
results obtained with the single particle and DCP wave functions. 
We see that, while both the FRDWBA and the AD model calculations reproduce
the peak value of the data \cite{nak1} well, the FRDWBA calculations done
with the DCP wave function overestimate it.  On the other hand, none of the
calculations is able to explain the data at higher relative energies. 
This can be attributed to the fact that nuclear breakup effects, which 
can contribute substantially \cite{dasso99} at higher relative energies
(for $E_{rel}$ $>$ 0.6 MeV), are not included in these calculations.
Of course, the authors of Ref. \cite{nak1} claim that their data have been
corrected for these  contributions. However, the procedure adopted
by them for this purpose is inadequate. They obtained the nuclear
breakup contribution on the Pb target, by scaling the cross sections
measured on a carbon target. This scaling procedure is unlikely to be
accurate for reactions induced by halo nuclei due to the presence of
a long tail in their ground state.
In a full quantum mechanical
theory, both Coulomb and nuclear breakup contributions should be calculated 
on the same footing and corresponding amplitudes should be added coherently
to get the cross sections. 

Calculations done using the BT theory (dashed line in the upper part
of Fig. 9) underestimates the data considerably. This difference between
the FRDWBA and the BT results can again be traced to the behavior
of $|F_r|^2$ in Fig. 4, which is smaller than unity at the beam energy
of 72 MeV/nucleon of this reaction.

We would like to remark that our post form FRDWBA results for the 
pure Coulomb breakup contribution to the relative energy spectra for the
$^{11}$Be agrees quite well with a recent calculation of the breakup
reaction in a non-perturbative approach where the time evolution
of the projectile is calculated by solving a time-dependent Schr\"odinger
equation \cite{mele99}. On the other hand, similar perturbative calculations
performed previously \cite{kid94} overestimate the data in the peak region.  

\subsubsection{Momentum distribution of the core}

The neutron halo structure is reflected in the narrow width of the parallel
momentum distribution (PMD) of the charged breakup fragments emitted 
in breakup reactions induced by the halo nuclei.
This is because the PMD has been found to be least affected
by the reaction mechanism \cite{orr95,mex,baz,bm92,ps95} and therefore, 
a narrow PMD can be related to a long tail in the  matter distribution
in the coordinate space via Heisenberg's uncertainty principle.
\begin{figure}[ht]
\begin{center}
\mbox{\epsfig{file=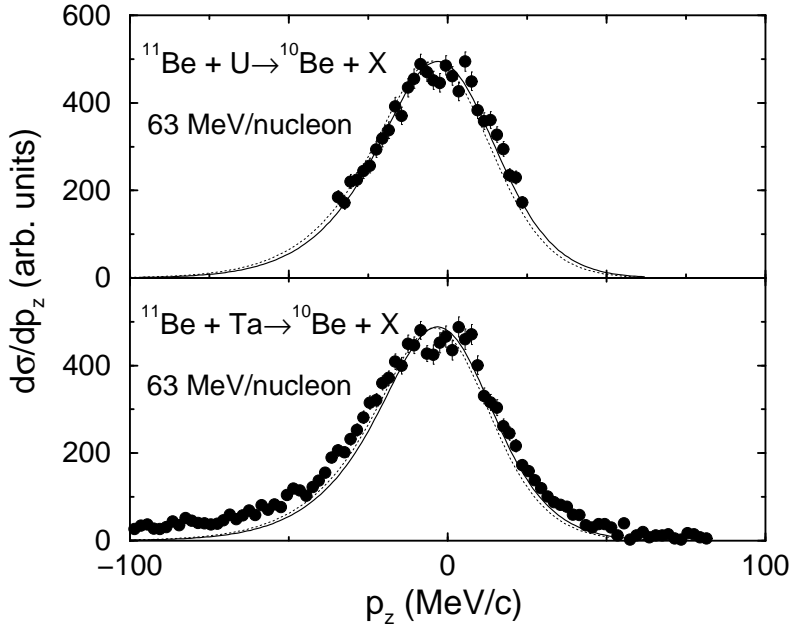,width=.75\textwidth}}
\end{center}
\caption{Parallel momentum distributions of $^{10}$Be in the breakup of 
$^{11}$Be on U (top half) and Ta (bottom half) at 63 MeV/nucleon beam energy 
in the rest frame of the projectile. The same normalization has been used for 
both the FRDWBA (solid line) and the AD (dotted line) cases. The data are
taken from \protect\cite{mex}.} 
\label{fig:figj}
\end{figure}

In Fig. 10 we present the PMD of the $^{10}$Be fragment emitted in the
breakup of $^{11}$Be on U and Ta targets at 63 MeV/nucleon beam 
energy. Calculations performed within both the FRDWBA and  
AD model formalisms using the configuration (a) are presented in this figure. 
The calculated cross sections are normalized to match 
the peak of the data points (which are given in 
arbitrary units) \cite{mex}, the normalization constant being the same 
for both cases. The full width at half maximum ( FWHM)
for the U and Ta targets are 44 MeV/c and 43 MeV/c respectively in both 
the FRDWBA and the AD cases. These agree well with the averaged 
experimental value of 43.6$\pm$1.1 MeV/c \cite{mex} and also with those 
calculated in \cite{pb1}. The very narrow widths of the parallel momentum
distributions signal the presence of a neutron halo structure in 
$^{11}$Be.

It may be noted that the calculations performed with configuration (c) gives 
results identical to that obtained with 
configuration (a). The PMD calculated with a pure $d$ -- wave configuration
is too small in magnitude and too wide in width. 
 
\subsubsection{Coulomb part of total one-neutron removal cross section}

For the breakup of $^{11}$Be on a Au target at the beam energy of
41 MeV/nucleon, 
the values of the pure Coulomb total one-neutron removal
cross section ($\sigma_{-n}^C$) are found to be 
1.91 b and 1.94 b for the FRDWBA and the AD model respectively, using 
configuration (a) and the single particle wave function for the
ground state of $^{11}$Be. The corresponding values of $\sigma_{-n}^C$ for
the breakup of this projectile on the Pb target at the beam energy of 
72 MeV/nucleon are 1.25 b and 1.29 barn respectively. 
The experimental values for the total one-neutron removal cross section
($\sigma_{-n}$) for these two  reactions are reported to be
2.5$\pm$0.5 b \cite{ann} and 1.8 $\pm$ 0.4 b \cite{nak1} respectively.
The difference between $\sigma_{-n}^C$ and the experimental value
of $\sigma_{-n}$ can be attributed to the nuclear breakup effects.
Incidentally, FRDWBA calculations performed with the DCP wave function
for the $^{11}$Be ground state leads to much larger values of 
2.82 b and 1.76 b for $\sigma_{-n}^C$ for the two cases respectively.  

\subsection{Results for $^{19,15,17}$C}

In this section,we shall compare results of our
calculations with the available data on the breakup of the neutron rich 
carbon nuclei $^{19,15,17}$C on heavier targets.

There is a large uncertainty in the value of the last neutron separation
energy in $^{19}$C ($S_{n-^{18}C}$) with quoted values varying
between 160 -- 530 keV \cite{nak2,nub}. It has recently been
shown \cite{pb3} that most of the available data on the Coulomb
dissociation of $^{19}$C can be satisfactorily explained within
the adiabatic model of Coulomb breakup with the one-neutron separation
energy of 530 keV. The ground state spin-parity of $^{19}$C has been  
quoted as $1/2^+, 3/2^+$ and $5/2^+$ \cite{rid97}. 
The relativistic mean field (RMF) \cite{rmf} as well as shell model
calculations using Warburton-Brown effective interaction \cite{war} predict
the spin-parity of the ground state of this nucleus to be $1/2^+$.

We use single particle wave functions which are constructed
by assuming a Woods-Saxon interaction between the valence neutron and the
charged core whose depth (for fixed values of the radius and diffuseness 
parameters) is adjusted to reproduce the binding energies of the nuclei under
investigation. The valence neutron binding energies, searched  
potential depths ($V_{depth}$) and calculated rms radii of the projectile 
with different configurations for the ground state for each isotope are 
summarized in Table 1.

\begin{table}
\begin{center}
\caption[T1]{Searched depths of the  Woods-Saxon potential for given
projectile configurations and binding energies ($\epsilon$) 
with radius and diffuseness parameters of 1.5 $fm$ and
0.5 $fm$. The rms radius of the projectile is also shown for each case. }
\vspace{1.1cm}
\begin{tabular}{|c|c|c|c|c|}
\hline
Projectile & Projectile & $\epsilon$ & $V_{depth}$ & Projectile rms  \\
 &configuration& (\footnotesize{MeV})& (\footnotesize{MeV})& radius (\footnotesize{$fm$})  
 \\ \hline
$^{19}$C & $^{18}$C $(0^+)~~\otimes~~1s_{1/2}\nu$ & 0.530 & 49.77 & 3.19 \\
         & $^{18}$C $(0^+)~~\otimes~~0d_{5/2}\nu$ & 0.530 & 53.58 & 2.95 \\
         & $^{18}$C $(0^+)~~\otimes~~ 1s_{1/2}\nu$ & 0.160 & 47.43 & 3.66 \\
 & & & &     \\
$^{15}$C & $^{14}$C $(0^+)~~\otimes~~ 1s_{1/2}\nu$ & 1.2181 & 61.21 & 2.83 \\
         & $^{14}$C $(0^+)~~\otimes~~ 0d_{5/2}\nu$ & 1.2181 & 65.66 & 2.67 \\
   & & & & \\
$^{17}$C & $^{16}$C $(0^+)~~\otimes~~ 1s_{1/2}\nu$ & 0.729 & 54.45 & 3.03 \\
         & $^{16}$C $(0^+)~~\otimes~~ 0d_{5/2}\nu$ & 0.729 & 58.70 & 2.81 \\
         & $^{16}$C $(2^+)~~\otimes~~ 1s_{1/2}\nu$ & 2.5 & 62.55 & 2.79 \\
         & $^{16}$C $(2^+)~~\otimes~~ 0d_{5/2}\nu$ & 2.5 & 60.82 & 2.86 \\ \hline
\end{tabular}
\end{center}
\end{table}

In this work, we consider two situations: (i) different binding energies
(530 keV and 160 keV) with the same configuration for $^{19}$C ground state
($^{18}$C $(0^+)\otimes 1s_{1/2}\nu$), and (ii) 
different configurations ($^{18}$C $(0^+)\otimes 1s_{1/2}\nu$ and 
$^{18}$C $(0^+)\otimes 0d_{5/2}\nu$) for $^{19}$C ground state with the same
binding energy (530 keV).
We have considered single particle wave functions in all the 
cases, except for the
160 keV case where we have additionally considered a DCP wave function  
for the $s$ -- state \cite{len99}. It may be noted that for all the 
single particle wave functions, we have used a spectroscopic factor of 1.0.
\begin{figure}[ht]
\begin{center}
\mbox{\epsfig{file=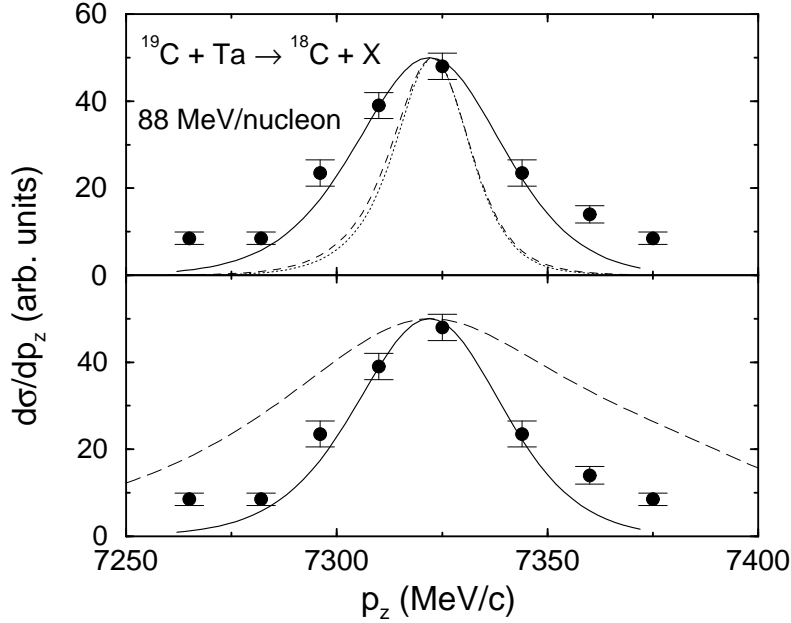,width=.75\textwidth}}
\end{center}
\caption{FRDWBA results for the parallel momentum distribution
of $^{18}$C in the breakup of $^{19}$C on Ta target at the beam energy
of 88 MeV/nucleon. The top half shows the results obtained with the configuration
($^{18}$C $(0^+)\otimes 1s_{1/2}\nu$) and single particle
wave function for the ground state of $^{19}$C with  
one-neutron separation energies of 530 keV (solid line), 160 keV (dashed line).
The dotted line shows the result obtained with a DCP wave function 
with a  one-neutron separation energy of 160 keV.
The bottom half shows the result obtained with the configurations
($^{18}$C$(0^+)\otimes 1s_{1/2}\nu$) (solid) and
($^{18}$C$(0^+)\otimes 0d_{5/2}\nu$) (dashed),  with the same value of the 
one-neutron separation energy (530 keV). The data have been
taken from \protect\cite{baz}.} 
\label{fig:figk}
\end{figure}
 
In Fig. 11, we present the PMD (calculated within the FRDWBA formalism) 
of the $^{18}$C fragment in the breakup of $^{19}$C on a Ta target at the
beam energy of 88 MeV/nucleon. We have normalized
the peaks of the calculated PMDs to that of the data (given in arbitrary units)
\cite{baz} (this also
involves coinciding the position of maxima of the calculated and experimental
PMDs). As can be seen from the upper part of this figure, the experimental
data clearly favor $S_{n-^{18}C}$ = 0.53 MeV with the 
$s$ -- wave n-$^{18}$C relative motion in the ground state of $^{19}$C. 
The results obtained with the $s$ -- wave configuration  within the
simple potential and DCP models (with the same value of $S_{n-^{18}C}$)
are similar to each other.

In the lower part of Fig. 11, we have shown the results obtained
with the $d$ -- wave relative motion for this system
(with $S_{n-^{18}C}$ = 0.53 MeV) and
have compared it with that obtained with a $s$ -- wave relative motion
with the same value of the binding energy. As can be seen,
the FWHM of the experimental PMD is grossly over-estimated by the $d$ -- wave
configuration.  The calculated FWHM with the $s$ -- state configuration
(with $S_{n-^{18}C}$ = 530 keV) is 40 MeV/c, which is in excellent agreement
with the experimental value of 41$\pm$3 MeV/c \cite{baz}. Thus these data
favor a configuration $^{18}$C($0^+)\otimes 1s_{1/2}$, with a one-neutron
separation energy of 0.530 MeV for the ground state of $^{19}$C. 
These results are in agreement with those of Ref. \cite{pb3}.
The narrow width of the PMD provides support to the presence of a 
one-neutron halo structure in $^{19}$C. 

\begin{figure}[ht]
\begin{center}
\mbox{\epsfig{file=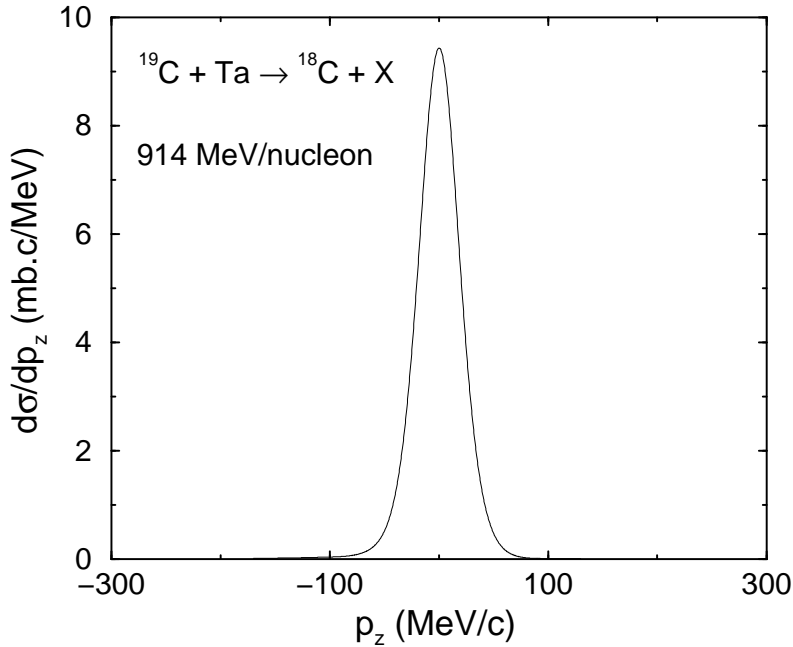,width=.75\textwidth}}
\end{center}
\caption{Parallel momentum distribution of $^{18}$C fragment in the Coulomb 
breakup of $^{19}$C on Ta target at the beam energy of 914 MeV/nucleon.}
\label{fig:figl}
\end{figure}

The width of the PMD is not expected to change with beam energy \cite{han95}.
To see this, we have calculated the PMD of $^{18}$C in the Coulomb
breakup of $^{19}$C on a heavy target (Ta) at the high beam energy of 914 
MeV/nucleon. The corresponding results are shown in Fig. 12. The 
FWHM of the distribution in this case is 42 MeV/c which is similar to that
obtained above at a lower beam energy. However, FWHM of the PMD of 
the $^{18}$C fragment measured in the breakup of $^{19}$C
on a carbon target  at the same beam energy, has been found \cite{bau98}
to be of 69$\pm$4 MeV/c.  
It would be interesting, therefore, to repeat measurement at this energy
with a heavier target to check our observation.
\begin{figure}[ht]
\begin{center}
\mbox{\epsfig{file=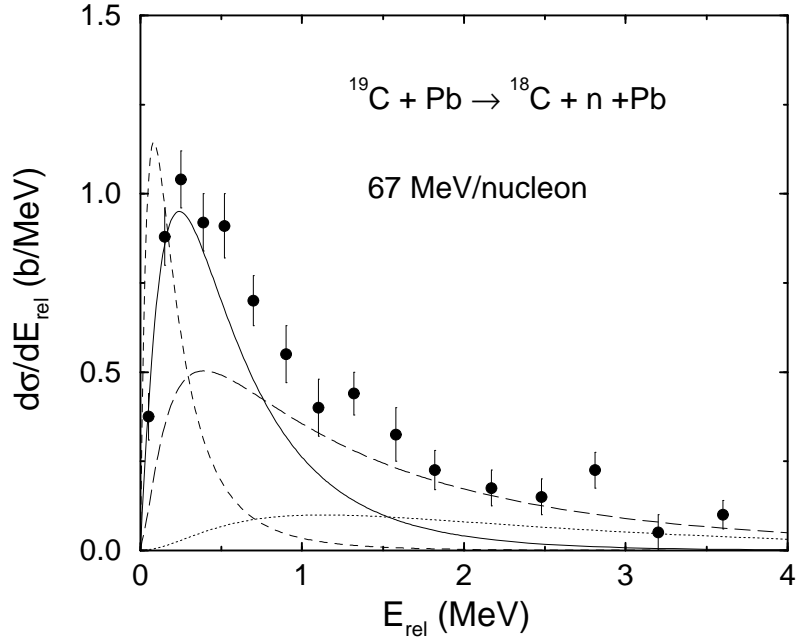,width=.75\textwidth}}
\end{center}
\caption{Calculated relative energy spectra for the Coulomb breakup of $^{19}$C
on Pb at 67 MeV/nucleon. The $d$ -- state results, for binding energies 160 keV 
(long dashed) and 530 keV (dotted), are multiplied by 10. The 160 keV 
$s$ -- state result (short dashed) is multiplied by 0.086. The 530 keV $s$ -- 
state result is represented by the solid line. The data have been
taken from \protect\cite{nak2}.} 
\label{fig:fign}
\end{figure}
 
In Fig. 13, we have shown the results of our FRDWBA calculations for the  
the relative energy spectrum for the 
breakup of $^{19}$C on Pb at 67 MeV/nucleon beam energy. The experimental
data is taken from \cite{nak2}. The angular integration for the $^{19}$C
center of mass is done up to the grazing angle of 2.5$^\circ$. It can be 
seen that in this case also the best agreement with the data (near the
peak position) is obtained with the $s$ -- wave configuration with
$S_{n-^{18}C}$ = 0.53 MeV. Calculations done 
with the $d$ -- state configuration for both 530 
keV and 160 keV one-neutron separation energy fails to reproduce
the data. At the same time, those performed with the $s$ -- state
configuration but with $S_{n-^{18}C}$ = 0.16 MeV overestimates the data 
by at least an order of magnitude and also fail to reproduce the
its peak position. However, as in the case of $^{11}$Be, our calculations
underestimate the relative energy spectrum for larger values of relative
energies. In this case also, the proper consideration 
of the nuclear breakup effects is
necessary to explain the data in this region. We would also
like to remark that the correction for these effects made in the data
\cite{nak2} by scaling the cross sections for the breakup of $^{19}$C on
a carbon target is unlikely to be accurate 
due to the same reasons as stated in case of $^{11}$Be.
\begin{table}
\begin{center}
\caption[T2]{Calculated values of $\sigma_{-n}^C$ for $^{19}$C.
$\epsilon$ is the one-neutron separation energy (MeV). The beam energy
($E_{beam}$) is 
in MeV/nucleon. The single particle wave functions are used in all
the cases.}
\vspace{1.1cm}
\begin{tabular}{|c|c|c|c|cc|cc|}
\hline
Projectile & $\epsilon$ & $E_{beam}$ & SF &
 \multicolumn{2}{|c|}{$\sigma_{-n}^C$ ($s$ -- state) (mb)} & 
\multicolumn{2}{|c|}{$\sigma_{-n}^C$ ($d$ -- state) (mb)} \\ \cline{5-8}
+ target& & & & FRDWBA & AD & 
FRDWBA & AD  \\ \hline

$^{19}$C~+~Ta & 0.530 & 88 & 1 & 780.4 & 780.9 & 62.5 & 66.8 \\
$^{19}$C~+~Ta & 0.160 & 88 & 1 & 4029.4 & 4072.0 & & \\
 & & & & & & & \\
$^{19}$C~+~Pb & 0.530 & 67 & 1 & 744.96 & & 25.7 & \\
$^{19}$C~+~Pb & 0.160 & 67 & 1 & 4246.0 & & 84.9 & \\ \hline
\end{tabular}
\end{center}
\end{table}

Table 2 summarizes the FRDWBA results of $\sigma_{-n}^C$ for the breakup of
$^{19}$C on different heavy targets at several beam energies.
The experimental value of $\sigma_{-n}$ for the breakup of $^{19}$C on
Ta at the beam energy of 88 MeV/nucleon is 1.1$\pm$0.4 b \cite{baz}.
It is seen from this
table that only with the $s$ -- wave configuration of the 
$^{19}$C ground state with $S_{n-^{18}C}$ = 0.53 MeV, the calculated
cross sections come closer to the experimental data. In this context it 
must be kept in mind that $\sigma_{-n}$ 
also include contributions from the nuclear breakup effects, which is
not included in our present calculations. 
 
We next consider the breakup of $^{15}$C which has a relatively larger
value for the one-neutron separation energy (1.2181 MeV)
and a ground state spin-parity of $1/2^+$ \cite{baz}. This can be
obtained from two configurations: a $1s_{1/2}$ 
neutron coupled to a $^{14}$C $(0^+)$ core and a $0d_{5/2}$ neutron 
coupled to a $^{14}$C $(0^+)$ core. One could also considered a    
$^{14}$C $(2^+)$ core and $0d_{5/2}$ neutron coupling to get a $1/2^+$ ground
state for $^{15}$C, but it would raise the one-neutron separation 
energy to about 7.01 MeV, which is highly unfavorable for the formation
of a halo. 
We, therefore, do not consider this configuration in our calculations.
 
\begin{figure}[ht]
\begin{center}
\mbox{\epsfig{file=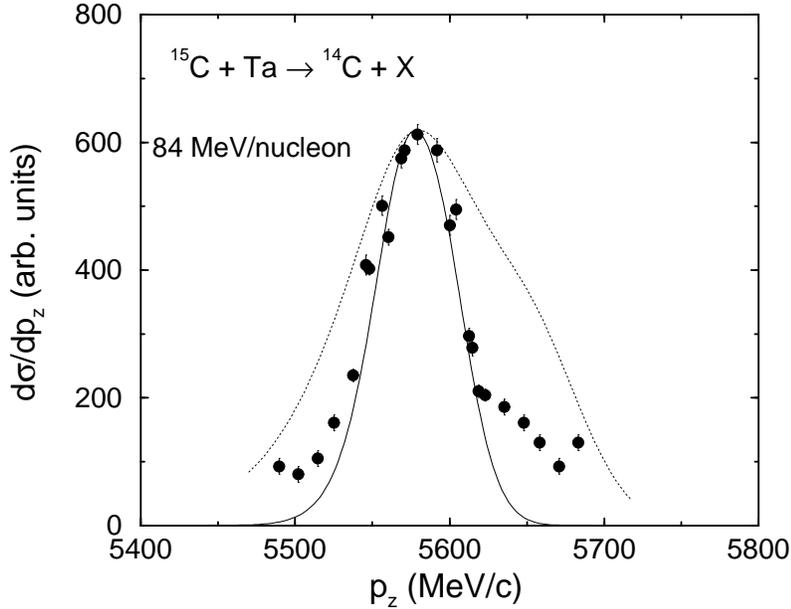,width=.75\textwidth}}
\end{center}
\caption{Parallel momentum distributions of $^{14}$C in the breakup of 
$^{15}$C on Ta at 84 MeV/nucleon. The solid line and dotted lines show the
results obtained with the configurations  
$^{14}$C $(0^+)\otimes 1s_{1/2}\nu$ 
the $^{14}$C $(0^+)\otimes 0d_{5/2}\nu$ respectively for the ground state of
the projectile. The data have been taken from \protect\cite{baz}.} 
\label{fig:figm}
\end{figure}

In Fig. 14, we present the results of our FRDWBA calculations for the PMD of
the $^{14}$C fragment in the breakup of $^{15}$C on the Ta target
at the beam energy of 84 MeV/nucleon. The experimental data is taken
from \cite{baz}. As before, 
our calculations are normalized to the peak of the data. 
The $s$ -- state configuration for the ground state of $^{15}$C gives a 
FWHM of 62 MeV/c, while with the $d$ -- state configuration it
comes out to be 140 MeV/c. Therefore, the experimental value for
the FWHM (67$\pm$1 MeV/c) \cite{baz} favors the former 
configuration. Hence our results provide support to the existence 
of a halo structure in $^{15}$C. This nucleus provides an example of  
the one-halo system with the largest one-neutron separation energy, 
known so far. 

$^{17}$C has a lower one-neutron separation energy (729 keV) as compared
to that of $^{15}$C. It would be interesting to see if it also has a 
halo structure, which seems probable if one considers only the binding
energies. The quoted ground state spin-parities for this nucleus are
$1/2^+, 3/2^+$ and $5/2^+$ \cite{rid97}. RMF calculations \cite{rmf}
predict it to have a value of $3/2^+$. We consider four possible ground
state configurations for this nucleus
and calculate the parallel momentum distributions of the $^{16}$C 
fragment in the breakup of $^{17}$C on a Ta target at 84 MeV/nucleon 
beam energy within our FRDWBA formalism. 
The FWHMs of the PMD obtained with different configurations are listed
in Table 3.

\begin{table}
\begin{center}
\caption[T3]{FWHMs from the parallel momentum distribution of $^{16}$C for
different ground state configurations of $^{17}$C and one-neutron
separation energies ($\epsilon$) in the breakup of $^{17}$C on Tai
at 84 MeV/nucleon beam energy.}
\vspace{1.1cm}
\begin{tabular}{|c|c|c|}
\hline
Projectile & $\epsilon$ & $ FWHM $   \\
configuration& (\footnotesize{MeV})& (\footnotesize{MeV/c})  
 \\ \hline
$^{16}$C $(0^+)~~\otimes~~ 1s_{1/2}\nu$ & 0.729 & 51 \\
$^{16}$C $(0^+)~~\otimes~~ 0d_{5/2}\nu$ & 0.729 & 114 \\
$^{16}$C $(2^+)~~\otimes~~ 1s_{1/2}\nu$ & 2.5 &  82\\
$^{16}$C $(2^+)~~\otimes~~ 0d_{5/2}\nu$ & 2.5 & 185 \\ \hline
\end{tabular}
\end{center}
\end{table}

It is evident from this table that the $s$ -- state configurations predict
a narrow width for the PMD, providing support to the existence
of a halo structure in this nucleus. The 
experimental data \cite{baz}, however, is available only for the breakup of 
$^{17}$C on a light target (Be) at 84 MeV/nucleon, which gives a FWHM of 
145$\pm$5 MeV/c. Since the PMD is mostly unaffected by the reaction
mechanism \cite{pb4}, it is quite likely that the experimental
FWHM will be the same also for the breakup of this nucleus on a 
heavier target. Therefore, the results shown in  
Table 3 seem to provide support to a $d$ -- wave configuration for the
ground state of $^{17}$C \cite{ang99}. Hence, the existence of a one-neutron
halo structure is quite improbable in $^{17}$C. However, to arrive at a more
definite conclusion in this regard the data on the breakup of   
$^{17}$C on a heavy target is quite desirable. 

\section{Summary and conclusions}

In this paper, we have performed calculations for the Coulomb breakup
of the neutron rich  nuclei $^{11}$Be and $^{15,17,19}$C, which
have a single valence neutron loosely bound with a stable core. We used
a theory developed within the framework of the post form distorted 
wave Born approximation where finite range effects have been included
approximately by using a local momentum approximation on the Coulomb
distorted wave of the outgoing charged fragment. Within this method,
the breakup amplitude is expressed
as a product of factors describing separately the projectile 
structure and the dynamics of the reaction. This factored form of the breakup
amplitude can also be obtained within an adiabatic model which makes the 
approximation that the strongly excited core-valence particle relative
energies in the Coulomb breakup are small. However, unlike the post form
DWBA, the adiabatic model does not use the weak coupling approximation 
to describe the center of mass motion of the fragments with respect to
the target. 

Both these theories allow the use of realistic wave functions
for the ground state of the projectile. Furthermore, unlike the 
semiclassical and quantum mechanical theories using the zero range
approximation which can be applied only to $s$ -- wave projectiles, these
methods are applicable to projectiles with any relative orbital angular
momentum structure between their fragments. This provides an opportunity
to probe the structure of the ground state of the projectile, by 
comparing the predictions of these theories with the data for the 
breakup observable. We have calculated, the energy, angular and
parallel momentum distributions of the fragments emitted in the breakup
reaction of these nuclei using different configurations for their
ground state, By  making comparisons of the calculated cross sections 
with the available experimental
data an effort has been made to put constraints on their ground state
structure.

All the observable calculated by us are sensitive to the ground state
configuration of the projectile. We find that for $^{11}$Be,
a $s$ -- wave configuration
($^{10}$Be$(0^+) \otimes 1s_{1/2}\nu$), with a 
spectroscopic factor of 0.74 for its ground state 
provides best agreement with the experimental data in all the cases. 
In our study, it is not possible to distinguish
between this configuration and the one proposed recently where 
there is an admixture of the $s$ -- wave and a $d$ -- wave configuration, 
($^{10}$Be$(2^+)\otimes 0d_{5/2}\nu$), 
with spectroscopic factors of 0.74 and 0.17 respectively 
for the ground state of this nucleus. For almost 
all the observables, there is a general agreement between the FRDWBA and
adiabatic model results even in the absolute magnitude in the region
where Coulomb breakup is expected to be dominant mode (ie. below
the grazing angle). This provides additional support to our choice of
the parameters associated with the 
local momentum in our FRDWBA calculations.
It may be noted that the approximation of 
Baur and Trautmann, which also leads to the factored form for the
breakup amplitude, gives results which are very different from those
obtained with the FRDWBA and AD model formalisms. The BT approximation fails
to explain the data in most of the cases studied here.

In the case of neutron angular distributions for 
the breakup of $^{11}$Be on the gold target at the beam energy of
41 MeV, there is a dip around 25$^\circ$ in the adiabatic 
model calculations, which is not seen in 
the corresponding FRDWBA results. It may be 
noted that this region was excluded in the results shown in Ref.
\cite{pb1}. This dip can be traced back to the fact that the form factor
in the adiabatic model has a node at the momentum transfer corresponding
to this angle. In any case, this region of the angular distribution is
expected to get substantial contribution from the nuclear breakup effects
(and also from the Coulomb-nuclear interference terms). Therefore, full
implication of this dip can become clear only after nuclear breakup effects
are included in these models. 

For the $^{19}$C case, the results for the PMD of $^{18}$C 
and the relative energy spectrum of the $n$ + $^{18}$C system
show that the most probable ground state configuration of $^{19}$C is
($^{18}$C $(0^+)\otimes 1s_{1/2}\nu$) with a one-neutron separation
energy of 530 keV and a spectroscopic factor of 1. Our FRDWBA
calculations agree
well with the experimental data of the MSU group \cite{baz}.
By performing the calculations at GSI energies we note that the width of the
PMD is independent of the beam energy. It would be interesting to perform 
the GSI experiment with a heavier target to check this prediction.
Our FRDWBA results on $^{19}$C are in excellent
agreement with those of the adiabatic model \cite{pb3}.

We find that the most probable configuration for
$^{15}$C is a $s$ -- wave valence
neutron coupled to the $^{14}$C core and that for $^{17}$C is a 
$d$ -- wave valence
neutron coupled to a $^{16}$C core. Both the 
experimental and the calculated FWHM of the
PMD for the $^{14}$C core in the breakup of $^{15}$C are small and they agree 
well with each other. This provides support to the existence of a one-neutron
halo structure in $^{15}$C. On the other hand, in the case of $^{17}$C the 
value of this quantity for the $^{16}$C core is probably closer to that of
a stable isotope. Therefore the existence of a halo structure in
$^{17}$C appears to be unlikely.  
Interestingly the one-neutron
separation energies of $^{15}$C and $^{17}$C are 1.2181 and 0.729 MeV 
respectively. So the binding energy
of the valence neutron as well as its configuration with respect to
the core together decide
whether a nucleus has halo properties or not. 

\section{ACKNOWLEDGMENTS}

The authors are thankful to Horst Lenske for providing them the wave functions
of the dynamical polarization model for various projectiles. One of the
authors (RS) wishes to acknowledge several fruitful discussions with Dr.
M. A. Nagarajan on the local momentum approximation method.

\appendix
\section{ Validity of the local momentum approximation} 
\begin{figure}[ht]
\begin{center}
\mbox{\epsfig{file=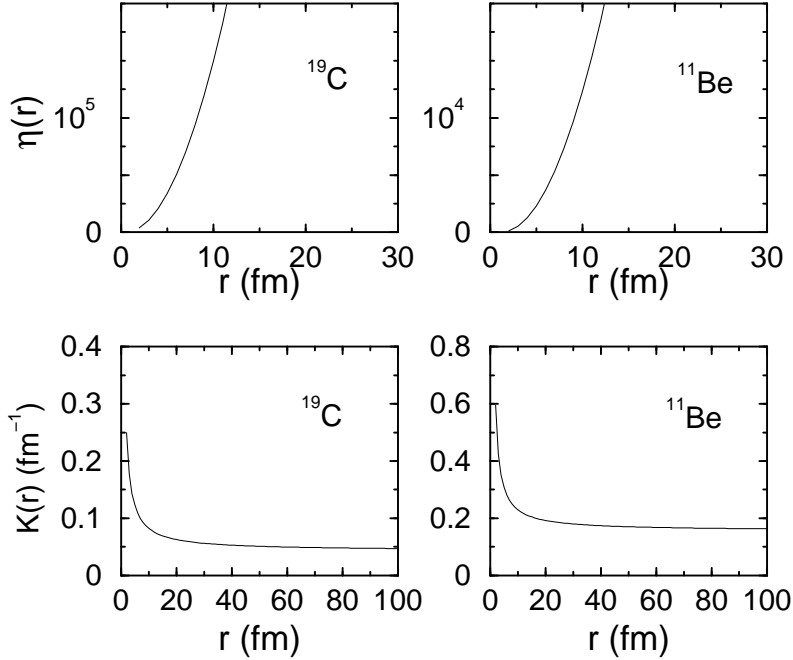,width=.75\textwidth}}
\end{center}
\caption{variation of $\eta (r)$ (upper half) and ${ K}(r)$
(lower half) with $r$.
}
\label{fig:figo}
\end{figure}

As discussed in \cite{shyam85}, a condition for the validity of the
local momentum approximation is that the quantity 
\begin{eqnarray}
\eta(r) = {1\over 2}{ K}(r)|d{ K}(r)/dr|^{-1}
\end{eqnarray}
evaluated at a representative distance $R$ should  
be much larger than ${\it R}_{bc}$, which is of the order of the range
of the interaction $V_{bc}$. To check this, we show in Fig. A.1, 
$\eta(r)$ (upper half) and ${ K}(r)$ (the magnitude of the local
momentum) (lower half) as a function of $r$, for the breakup reactions
$^{19}$C + Ta $\rightarrow$ $^{18}$C + $n$ + Ta at the beam energy of  
88 MeV/nucleon (left side) and  
$^{11}$Be + Au $\rightarrow$ $^{10}$Be + $n$ + Au at the beam energy of
41 MeV/nucleon (right side). We see that for $r$ $>$ 8 $fm$,
$\eta(r)$ is several orders of magnitude larger than rms radii of the
halo in both the cases. Therefore, the above condition is well satisfied.  

From the lower half of Fig. A.1, we note that the value of ${ K}(r)$
remains constant for distances larger than 10 $fm$. Due to the peripheral
nature of the breakup reaction, this region contributes maximum to the 
cross section. Therefore, our choice of a constant  magnitude for
the local momentum evaluated at 10 $fm$ is well justified. In fact, we noted 
that as $R$ is increased from 5 to 10 $fm$ the calculated cross sections
vary by at the most 10$\%$, and with a further increase the variation
is less than 1$\%$, in all the cases considered in this paper. 
\begin{figure}[ht]
\begin{center}
\mbox{\epsfig{file=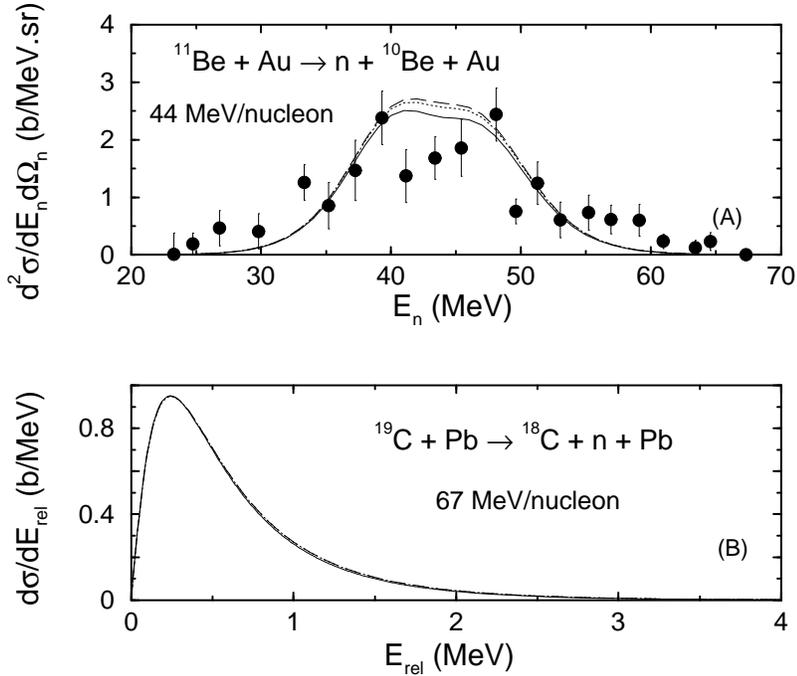,width=.75\textwidth}}
\end{center}
\caption{(A) The energy distributions of neutron  
from the breakup of $^{11}$Be on 
Au at the beam energy of 41 MeV/nucleon and (B) relative energy spectra
for the dissociation of $^{19}$C on Pb at 67 MeV/nucleon 
for choices $d_1$ (solid line), $d_2$ (dashed line) and $d_3$ (dotted line)
for the  direction of the local momentum, as discussed in the text. 
For the relative energy spectra the results obtained in three cases are
not distinguishable from each other. 
} 
\label{fig:figp}
\end{figure}

In the application of the LMA in the description 
of the heavy ion induced transfer reactions, it was noted \cite{braun74b}
that the calculated cross sections were more or less unaffected by the
choice of the direction of the local momentum. However,
in Ref. \cite{pb1}, some dependence of the breakup cross sections on this
direction has been reported. To study the sensitivity of our FRDWBA results 
on the direction of ${ {\bf K}}$, we have performed calculations for
three cases where we take the angles of the local momentum
($d_1$) parallel to those of ${\bf k}_b$, ($d_2$) parallel to the direction
corresponding to the half of the angles of ${\bf k}_b$ and ($d_3$) parallel
to the beam direction (zero angles).

In Table A.1, we show the results for $\sigma_{-n}^C$ for the breakup
of $^{11}$Be and $^{19}$C (on  targets and at beam energies as indicated
therein), for these three choices of the direction of ${ {\bf K}}$.
We see that between the cases $d_1$ to $d_3$,
the variation in the values of $\sigma_{-n}^C$ is less than  5$\%$
for $^{11}$Be, and less than 2$\%$ for $^{19}$C. 
In part A of Fig. A.2, we show the energy distribution
of the neutron for the same reaction as described in Fig. 2  
The results obtained with cases $d_1$, $d_2$ and $d_3$ are shown
by solid, dashed and dotted lines respectively. We note that energy
distributions calculated with these choices differ from each other
only in the peak region; the variation between them is of the order
of only 5$\%$. The results for the neutron angular distribution
(for the reaction reported in Fig. 8) is of the similar nature.

In part B of Fig. A.2, we show the relative energy spectrum for the
same reaction as shown in Fig. (13), for the choices ($d_1$), ($d_2$)
and ($d_3$). In this case we observe almost no variation in the
calculated cross sections.  Similarly, We have noted no dependence
of the calculated widths of the  parallel momentum distributions of heavy
fragments on the direction of ${ {\bf K}}$, in all the reactions 
investigated in this paper. Therefore, the dependence of various cross sections 
for the reaction studied in this paper, on the direction of the
local momentum is either very minor or almost negligible. The measurements
done so far are not able to distinguish the small differences that we see
here in some cases. Therefore, we have 
performed all our calculations in this paper by using 
${\hat {{\bf K}}}$ = ${\hat {\bf k}}_b$.

\begin{table}
\begin{center}
\caption[T4]{Calculated value of the Coulomb part of one-neutron
removal cross section for $^{11}$Be
for three different directions of local momentum ($d_1$) 
($d_2$) and ($d_3$). }
\vspace{1.1cm}
\begin{tabular}{|ccccccc|}
\hline
Projectile & $\epsilon$ & $E_{beam}$& SF &  $\sigma_{-n}^C$& $\sigma_{-n}^C$&
$\sigma_{-n}^C$\\
+ target& (\footnotesize{MeV})& (\footnotesize{MeV/nucleon})& & $d_1$& $d_2$ & 
$d_3$\\
       &                     &                             & & (mb) & (mb) & 
(mb) 
\\ \hline
$^{11}$Be~+~Au & 0.504 & 41 & 0.74 & 1906 & 1973 & 1979  \\ 
$^{19}$C~+~Pb & 0.530 & 67 & 1 & 745 & 758 & 760 \\ \hline
\end{tabular}
\end{center}
\end{table}

\end{document}